\documentclass[prd,10pt,aps,preprintnumbers,twocolumn,superscriptaddress,showpacs,tightenlines,floatfix,amssymb]{revtex4-1}
\usepackage{epsfig}

\newcommand{\beqn}{\begin{eqnarray}}
\newcommand{\eeqn}{\end{eqnarray}}
\newcommand{\eq}[1]{(\ref{#1})}

\newcommand{\dd}{\!{\mathrm d}}

\begin{document}

\title{Pipelike current-carrying vortices in two-component condensates}

\author{M. N. Chernodub}\thanks{On leave from ITEP, Moscow, Russia.}
\affiliation{Laboratoire de Math\'ematiques et Physique Th\'eorique,
Universit\'e Fran\c{c}ois-Rabelais Tours,
F\'ed\'eration Denis Poisson - CNRS,
Parc de Grandmont, 37200 Tours, France}
\affiliation{Department of Physics and Astronomy, University of Gent, Krijgslaan 281, S9, B-9000 Gent, Belgium}
\author{A. S. Nedelin}
\affiliation{Department of Physics and Astronomy, Uppsala University, P.O. Box 803, Uppsala, S-75108, Sweden}
\affiliation{Institute for Theoretical and Experimental Physics, B.~Cheremushkinskaya 25, Moscow, 117218, Russia}

\begin{abstract}
We study straight vortices with global longitudinal currents in the Bogomolny limit of the Abelian Higgs model with
two charged scalar fields. The model possesses global $SU(2)$ and local electromagnetic $U(1)$ symmetries
spontaneously broken to global $U(1)$ group, and corresponds to a semilocal limit of the standard electroweak model.
We show that the contribution of the global $SU(2)$ current to the vortex energy
is proportional to the total current squared. Locally, these vortices carry also longitudinal electromagnetic currents,
while the total electromagnetic current flowing through a transverse section of the vortex is always zero.
The vortices with high winding numbers have, in general, a nested pipelike structure. The magnetic field of the vortex
is concentrated at a certain distance from the geometric center of the vortex, thus resembling a ``pipe.'' This magnetic
pipe is layered between two electrically charged pipes that carry longitudinal electric currents in opposite directions.
\end{abstract}


\preprint{UUITP-20/10}

\date{\today}

\maketitle

\section{Introduction}

Certain extended topological defects are able to support longitudinal currents. A well-known example of such a defect
is the Abrikosov--Nielsen--Olesen vortex solution~\cite{ref:ANO} in a $U(1) \times U(1)$ model~\cite{Witten:1984eb}:
the corresponding vortex string may carry both bosonic and fermionic superconducting flows of large magnitude. Loops
of these superconducting strings may have various cosmological consequences~\cite{Ostriker:1986xc}.

The extended topological structures with longitudinal currents appear also at much smaller scales in the context of
condensed matter physics.  For example, vortices that carry longitudinal superfluid flow in a superfluid ${}^3$He-A were
proposed theoretically in Ref.~\cite{ref:GEV1}. Later, certain signatures of these stringy objects,
known as $w$ vortices, were found in nuclear magnetic resonance experiments~\cite{ref:GEV2}
(see Ref.~\cite{ref:Volovik:review} for a detailed review).

The symmetry patterns and the zoo of the topological defects in the superfluid ${}^3$He have a lot in common with
the corresponding properties of the standard electroweak model in particle physics~\cite{Volovik:2003fe}. Thus it is not
surprising that one of the most interesting realizations of the current-carrying strings in the field theory context appears in
the standard model of electroweak interactions. This model supports various (embedded) topological defects including well-known
$Z$- and $W$ vortices~\cite{Vachaspati:1991dz,Hindmarsh:1991jq}.
In a  semilocal limit of the model  (characterized by a special value of the Weinberg angle, $\theta_W = \pi/2$) these
vortex solutions are known to carry persistent longitudinal currents of isocharge associated with the global symmetry subgroup~\cite{Forgacs:2005sf}.
The word ``semilocal'' indicates that the symmetry group of the model is a product of the local electromagnetic group, $U_{e.m.}(1)$, and the global
isosymmetry group, $SU_I(2)$. In this limit the original non-Abelian {\it gauge} group decouples from the scalar fields, and the resulting theory
is basically the Abelian Higgs model with two complex scalar fields.

Beyond the semilocal limit (i.e., at any value of Weinberg's angle)
the current-carrying solutions were constructed
in Refs.~\cite{Volkov:2006ug,Garaud:2009uy} and evidence of their perturbative stability was found~\cite{Garaud:2009uy}.
We refer an interested reader to the reviews~\cite{Achucarro:1999it} and \cite{Radu:2008pp} for an extended discussion of
the vortex solutions in the standard electroweak model.

In the semilocal limit the vortices with nonzero longitudinal isocurrents exist only in type--II regime~\cite{Forgacs:2005sf}
because the energy per unit vortex length of such vortices diverges as the system approaches the boundary between type-II and
type-I superconductivity. This boundary is known as the Bogomolny limit~\cite{ref:Bogomolny}. General vortex solutions
without longitudinal currents were described in the Bogomolny limit of the semilocal model in Ref.~\cite{Gibbons:1992gt}.
The (iso)charged vortex solutions with persistent longitudinal currents were found in the same limit in Ref.~\cite{Abraham:1992hv}.
We continue this line of investigation by studying in details the current-carrying vortex solutions in the Bogomolny limit of the
Abelian with two Higgs fields. We concentrate on the current-carrying vortices with multiple winding numbers because these vortices
-- contrary to the elementary vortices with a unit winding number~\cite{Abraham:1992hv} -- do have a finite energy per
unit vortex length.

Our study is also motivated by the (theoretically anticipated and/or experimentally found) existence of various systems with multiple-component
order parameters such as the already mentioned superfluid ${}^3$He~\cite{ref:Volovik:review},
two-component Bose-Einstein condensates~\cite{ref:Matthews},
two-band superconductors that includes the well-studied system of MgB${}_2$~\cite{ref:MgB2:1},
liquid metallic hydrogen~\cite{ref:Egor}, two-component plasmas~\cite{Faddeev:2000rp} etc.

The structure of this paper is as follows. In Sec.~\ref{sec:two-component} we describe the Abelian model with two
charged scalar complex fields. This model admits vortex solutions to the classical equation of motion, which
are drastically simplified in the Bogomolny limit. Section~\ref{sec:pipelike} is devoted to the illustration of various
multivortex solutions in this limit. We concentrate on the vortices with high winding numbers which have a
``pipelike'' structure: the magnetic field in such vortices is nonzero only in a thin region located at a certain fixed
distance from the geometrical center of the vortex. In Sec.~\ref{sec:longitudinal} we introduce a longitudinal
wave along the vortex and check that this wave corresponds to a current-carrying solution that satisfies the original (second-order)
classical equations of motion. Finally, in Sec.~\ref{sec:energy} we calculate the energy per unit length of these vortices.
The last section is devoted to our conclusion.

\section{Two-component model}
\label{sec:two-component}

The Lagrangian of the two-component Ginzburg-Landau model is
\begin{equation}
{\cal L} = -\frac{1}{4}F_{\mu\nu} F^{\mu\nu} + \frac{1}{2}\left|D_{\mu}\Phi\right|^{2} - \lambda\left(|\Phi|^{2}-\eta^{2}\right)^{2}\,,
\label{lagrange}
\end{equation}
where $F_{\mu\nu} = \partial_\mu A_\nu - \partial_\nu A_\mu$ is the field strength tensor of
the $U_{e.m.}(1)$ gauge field $A_\mu$ and $D_\mu = \partial_\mu + i e A_\mu$ is the corresponding covariant derivative,
and $|\Phi|^2 \equiv \rho^2 = \Phi^\dagger \Phi$. The scalar field $\Phi$ has two complex-valued components,
\beqn
\Phi =
\left(
\begin{array}{c}
\rho_1 \, e^{i \varphi_1} \\
\rho_2 \, e^{i \varphi_2}
\end{array}
\right)
\equiv \rho
\left(
\begin{array}{c}
e^{i \varphi_1} \cos \theta \\
e^{i \varphi_2} \sin \theta
\end{array}
\right)
\,,
\label{eq:Phi}
\eeqn
where we introduced the parametrization in terms of the
two phases $\varphi_{1,2} \in [-\pi,\pi)$, two condensates $\rho_{1,2}$
with
\beqn
\rho = \sqrt{\rho_1^2 + \rho_2^2}\,,
\label{eq:rho12}
\eeqn
and/or the angular variable $\theta \in [0,\pi)$.

The model~\eq{lagrange} possess the symmetry
\begin{equation}
G = SU_I(2)\times U_{e.m.}(1)\,,
\label{eq:G}
\end{equation}
where $SU_I(2)$ group rotates the Higgs fields in the isotopic space,
\begin{equation}
SU_I(2): \quad \left\{
\begin{array}{l}
\Phi \to U \Phi\,, \\
A_\mu \to A_\mu \,,
\end{array}
\right.
\label{eq:SU2I}
\end{equation}
with $U \in SU_I(2)$, and the Abelian $U_{e.m.}(1)$ group corresponds to the local electromagnetic symmetry
which affects the phases of the scalar field~\eq{eq:Phi},
\begin{equation}
U_{e.m.}(1): \quad
\quad \left\{
\begin{array}{l}
\Phi \to e^{i \gamma(x)} \Phi\,, \\
A_\mu \to A_\mu - \frac{1}{e} \partial_\mu \gamma(x)\,.
\end{array}
\right.
\label{eq:Uem}
\end{equation}
According to Eq.~\eq{eq:Uem} both components of the scalar field~\eq{eq:Phi} carry the same electric charge $e$.

The dimensionality of the symmetry group~\eq{eq:G} indicates that there are four conserved Noether currents.
The electric (super)current corresponds to the rotations in the $U_{e.m.}(1)$ direction,
\beqn
J_\mu \equiv \rho^2 j_\mu = \frac{1}{2 i} [\Phi^\dagger D_\mu \Phi - c.c.]\,.
\label{eq:J}
\eeqn
For the sake of further convenience we introduced here the reduced electric current:
\beqn
j_\mu = J_\mu / \rho^2\,.
\label{eq:reduced}
\eeqn

The three other Noether currents can be grouped into the isovector field
\beqn
\vec{K}_{\mu} & = & 2eJ_{\mu}\vec{n}-\rho^{2}\vec{n}\times\partial_{\mu}\vec{n}\,,
\label{eq:K}\\
\vec{n} & = & \frac{\Phi^{\dagger}\vec{\tau}\Phi}{\Phi^{\dagger}\Phi}\,, \qquad \vec{n}^2 = 1\,.
\label{eq:n}
\eeqn
The isocurrent $\vec{K}_{\mu}$ corresponds to the isospinor rotations $SU_I(2)$.
The unit isovector $\vec{n}$ -- that is constructed with the Pauli matrices
$\vec\tau = (\tau^1,\tau^2,\tau^3)$ -- transforms in adjoint
representation of the $SU_I(2)$ subgroup~\eq{eq:SU2I}. The Abelian gauge transformations~\eq{eq:Uem}
leave both the isocurrent \eq{eq:K} and the vector~\eq{eq:n} intact.

It is convenient to make the rescaling,
\begin{eqnarray}
\Phi\to\eta\Phi,
\quad
A_{\mu}\to\eta A_{\mu},
\quad
x_{\mu}\to\frac{x_{\mu}}{e\eta},
\label{eq:rescaling}
\end{eqnarray}
so that all variables appear to be dimensionless. Then the
classical equations of motion of the model~\eq{lagrange} are
\begin{eqnarray}
\partial^{\mu}j_{\mu\nu}+\rho^{2} j_{\nu} - \frac{1}{2}\partial^{\mu}f_{\mu\nu} & = & 0\,,
\quad\label{eq1}\\
-\partial^{2}\rho + \frac{1}{4} \rho \, \partial_{\mu}\vec{n} \, \partial^{\mu}\vec{n}
+ \rho j_{\mu} j^{\mu} - 4 \alpha \rho\left(\rho^{2}-1\right) & = & 0\,,
\quad\label{eq2}
\end{eqnarray}
where we denoted $\alpha = \lambda/e^2$, and
\begin{eqnarray}
j_{\mu\nu} & = & \partial_{\mu}j_{\nu}-\partial_{\nu}j_{\mu}\,,
\label{eq:fj1}\\
f_{\mu\nu} & = & \vec{n}\cdot\partial_{\mu}\vec{n}\times\partial_{\nu}\vec{n}\,,
\label{eq:fj2}
\end{eqnarray}
are the strength tensors for the (reduced) electric current $j_\mu$, Eq.~\eq{eq:reduced}, and for the vector field $\vec n$,
Eq.~\eq{eq:n}, respectively.
Instead of the gauge-variant field $A_\mu$ we use below the gauge-invariant electric current $j_\mu$ as an independent variable.
The change from gauge-variant variables $A_\mu$ and $\Phi$ to the gauge-invariant fields $j_\mu$, $\vec n$ and $\rho$, is
commonly used in this model~\cite{Babaev:2001zy,ref:Govaerts:2001}.

Equations~\eq{eq1} and \eq{eq2} are supplemented by the conservation laws of the electric current \eq{eq:J}
and the isocurrent~\eq{eq:K}, respectively:
\begin{equation}
\partial_{\mu}J_{\mu} = 0\,, \qquad \partial_{\mu} \vec{K}_\mu = 0\,.
\end{equation}

\section{Pipelike vortex solutions in Bogomolny limit}
\label{sec:pipelike}

In this section we discuss known static straight vortexlike solutions to the classical equations of motion~\eq{eq1} and \eq{eq2}.
These vortices do not carry longitudinal currents so that the corresponding solutions depend only on two-dimensional
coordinates $x_1$ and $x_2$. Since all vectors/tensors involving (at least one of) the components $\mu=0,3$ are zero,
the system becomes effectively two-dimensional.

In a special limit,
\beqn
\alpha \equiv \frac{\lambda}{e^2} = \frac{1}{8}\,,
\label{eq:Bogomolny}
\eeqn
the classical equations of motion \eq{eq1} and~\eq{eq2} reduce to a simpler system of differential equations:
\begin{eqnarray}
\partial_{i}\vec{n}+\epsilon_{ij}[\vec{n}\times\partial_{j}\vec{n}] & = & 0\,,
\label{eq:bogomolny1}\\
\frac{1}{x}\frac{\partial}{\partial x}\left[x\frac{\partial}{\partial x}(\log\rho)\right]+\frac{1}{2}(\rho^{2}-1) - \frac{f_{12}}{2} & = & 0\,,
\label{eq:bogomolny2}
\end{eqnarray}
where $x = |\vec x|$.
The expressions for the electric current~\eq{eq:J} and the isocurrent~\eq{eq:K} are simplified as well:
\begin{equation}
J_i= - \epsilon_{ij} \rho \, \partial_j \rho\,,
\qquad
{\vec K}_i = - \epsilon_{ij} \partial_j (\rho^2 {\vec n})\,.
\label{eq:JK}
\end{equation}

The special choice of couplings~\eq{eq:Bogomolny} is known as the Bogomolny limit~\cite{ref:Bogomolny}.
The vortex solutions of the ``semilocal'' Abelian two-Higgs model -- that were first found in Ref.~\cite{Vachaspati:1991dz} --
were subsequently studied in this limit in Refs.~\cite{Hindmarsh:1991jq,Gibbons:1992gt}. A general solution to the classical
equations of motion in the Bogomolny limit~\eq{eq:Bogomolny} has a nonzero condensate~\eq{eq:rho12} at its origin and
in general is dubbed as ``skyrmion'' (for a detailed review see Ref.~\cite{Achucarro:1999it}).

In the Bogomolny limit the magnetic flux of the vortex is quantized:
\begin{eqnarray}
{\mathcal F} \equiv \oint_{C_\infty}  d x_i\, A_i = \int d^2 x \, B = \frac{2 \pi}{e} k\,, \qquad k \in \mathbb{Z}\,, \qquad
\label{eq:quant:F}
\end{eqnarray}
where
\begin{eqnarray}
B = F_{12} \equiv j_{12} - \frac{1}{2e} f_{12} = \frac{e}{2} (1 - \rho^2)\,,
\label{eq:B}
\end{eqnarray}
is the magnetic field of the vortex. The first integration in Eq.~\eq{eq:quant:F}
is taken over the spatial contour $C_\infty$ with an infinitely large radius. The last relation
in Eq.~\eq{eq:B} is supported by equations of motion~\eq{eq:bogomolny2} and Eq.~\eq{eq:JK}.

The string tension (i.e., the vortex energy per its unit length) is quantized as well (we restore $\eta$ for a moment):
\begin{eqnarray}
\sigma & = & \int d^2 x \left[ \frac {1}{2} B^2 + |D_i \Phi|^2 + \lambda\left(|\Phi|^{2}-\eta^{2}\right)^{2} \right]
\nonumber\\
& = & \frac{e \eta^2}{2} |{\mathcal F}| \equiv \pi \eta^2 |k|\,,
\label{eq:energy:functional}
\end{eqnarray}
where $i=1,2$ are two-dimensional indices in the transverse (with respect to the straight vortex) plane.

The validity of the quantization~\eq{eq:quant:F} can be checked in the angular pa\-ra\-metri\-za\-tion of
the scalar field~\eq{eq:Phi}. The energy functional~\eq{eq:energy:functional} contains the scalar derivative squared,
\begin{eqnarray}
 |D_i \Phi|^{2} & = & (\partial_\mu\rho)^{2} + \rho^{2} (\partial_i\theta)^{2} + J_i^{2}
\label{eq:Di}\\
               & & + \frac{\rho^2}{4}\sin^{2}2\theta \, [\partial_{\mu}(\varphi_{2} - \varphi_{1})]^{2}\,,
\nonumber
\end{eqnarray}
For the energy to be finite, a vacuum state is to be realized at the spatial infinity. This means that all four terms in
Eq.~\eq{eq:Di} should be zero: the variables $\theta$ and $\rho \neq 0$ should be independent of the coordinates, and
the electric current,
\beqn
J_i = \rho (e A_i+\cos^{2}\theta\partial_i\varphi_{1}+\sin^{2}\theta\partial_i\varphi_{2})\,,
\eeqn
should be vanishing at spatial infinity. Moreover, if at spatial infinity the angle $\theta$ does not take the specific values,
$\theta_\infty \neq 0$ and  $\theta_\infty \neq \pi/2$, then the two phases of the Higgs fields should be equal, $\varphi_1 = \varphi_2$.
We then recover the quantization of the magnetic flux \eq{eq:quant:F}, where the number $k$ gets interpreted as common winding number
of the phases $\varphi_1$ and $\varphi_2$.  If either $\theta_\infty = 0$ or $\theta_\infty = \pi/2$, then $\varphi_1 \neq \varphi_2$
and at large distances the winding number is carried either by $\varphi_1$ or by $\varphi_2$, respectively.

Using a stereographic projection one can parametrize the vector $\vec{n}$, Eq.~\eq{eq:n},
by a comp\-lex function $u = u(z)$ with $z = x_1 + ix_2$:
\begin{eqnarray}
n_{1}+in_{2}=\frac{2u}{1+|u|^{2}}\,, \qquad n_{3}=\frac{1-|u|^{2}}{1+|u|^{2}}\,.
\label{stereodef}
\end{eqnarray}
Then the first equation of motion~\eq{eq:bogomolny1} is satisfied automatically provided $u$ is a meromorphic function~\cite{Gibbons:1992gt}.

For a vortex located at the origin of the two-di\-men\-sion\-al plane, the following boundary conditions are to be implied:
\beqn
\lim_{z \to \infty} {\vec n} = (0,0,1)^T\,,
\qquad
\lim_{z \to 0} \vec{n} = (0,0,-1)^T\,,
\label{eq:BC}
\eeqn
where the first relation is the choice of the vacuum state. In terms of the function $u$ the conditions~\eq{eq:BC} read as
\begin{equation}
u(z=0)=\infty\,,\qquad u(z=\infty)=0\,.
\label{bound}
\end{equation}
The suitable meromorphic function is
\begin{equation}
u=\left(\frac{a}{z}\right)^{k},
\label{eq:sol}
\end{equation}
where $k$ is the mentioned winding number and
\begin{equation}
a = e \eta R_{\mathrm{vort}}
\label{eq:a:vort}
\end{equation}
is a dimensionless parameter  which defines the size $R_{\mathrm{vort}}$ of the vortex core.
This parameter is equivalent to the parameter $\xi_0$ of \cite{Hindmarsh:1991jq} and
to the parameter $|q_0|$ of \cite{Gibbons:1992gt}.

Substituting the solution~\eq{eq:sol} into Eq.~\eq{stereodef} and, subsequently, into
Eq.~\eq{eq:fj2}, one gets an expression for $f_{12}$. Then, introducing a normalized coordinate
\beqn
\xi = \frac{x}{a} \equiv \frac{r}{R_{\mathrm{vort}}}\,,
\label{eq:xi}
\eeqn
we rewrite the second equation of motion~\eq{eq:bogomolny2} as follows:
\begin{eqnarray}
\frac{1}{\xi}\frac{\partial}{\partial \xi}\left[\xi\frac{\partial}{\partial \xi}(\log\rho)\right]+\frac{a^2}{2}(\rho^{2}-1)
= \frac{2 k^2 \xi^{2 k - 2}}{(\xi^{2k} + 1)^2}\,.
\label{eq:rho}
\end{eqnarray}
This equation fixes the behavior of the scalar condensate $\rho$ as a function of distance from the vortex center $\xi$, Eq.~\eq{eq:xi};
for fixed vorticity $k$, Eq.~\eq{eq:quant:F}; and for the fixed size of the vortex core $a$, Eq.~\eq{eq:a:vort}.

\begin{figure}[!htb]
\begin{center}
  \includegraphics[width=55mm, angle=-90]{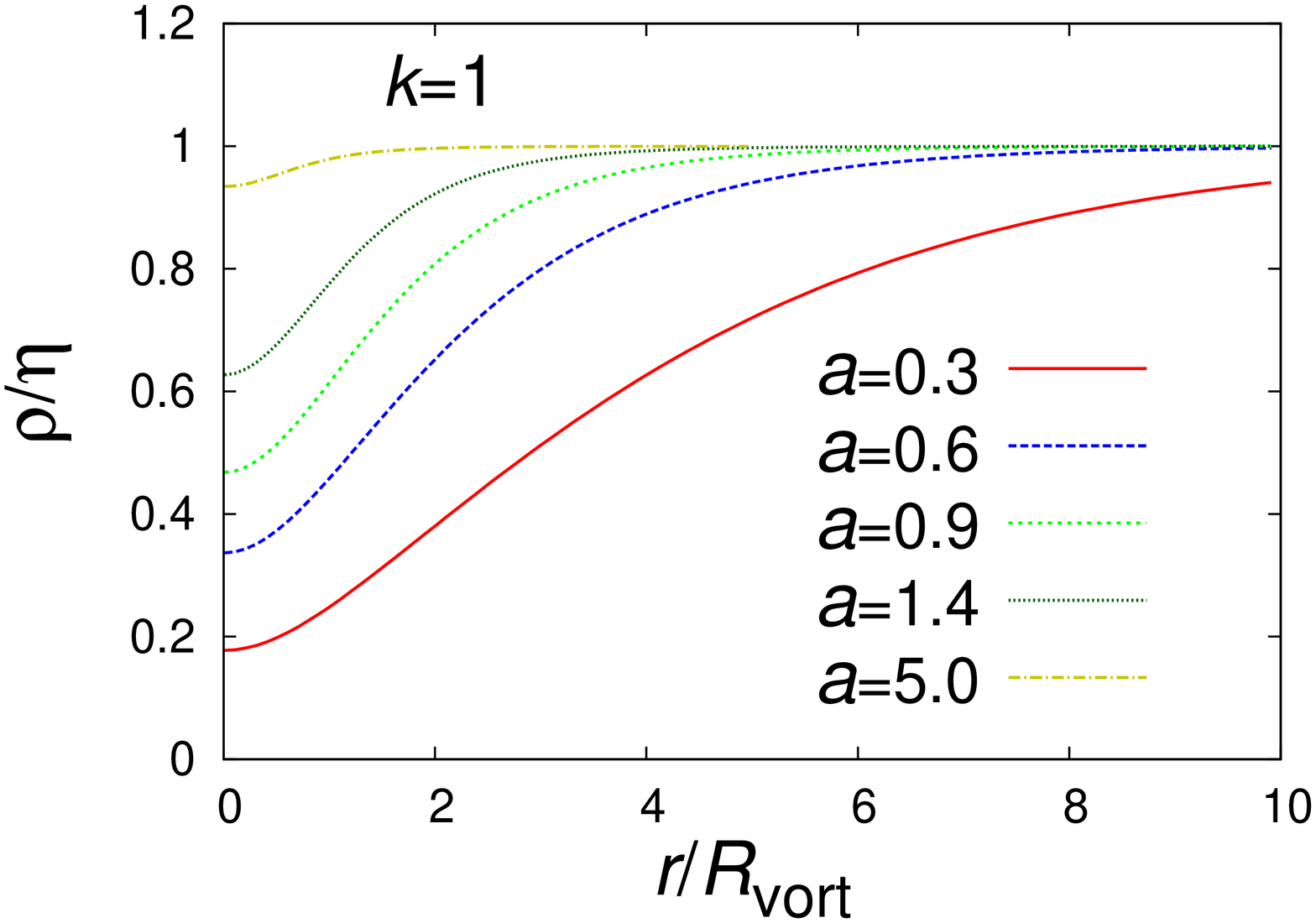}
  \includegraphics[width=55mm, angle=-90]{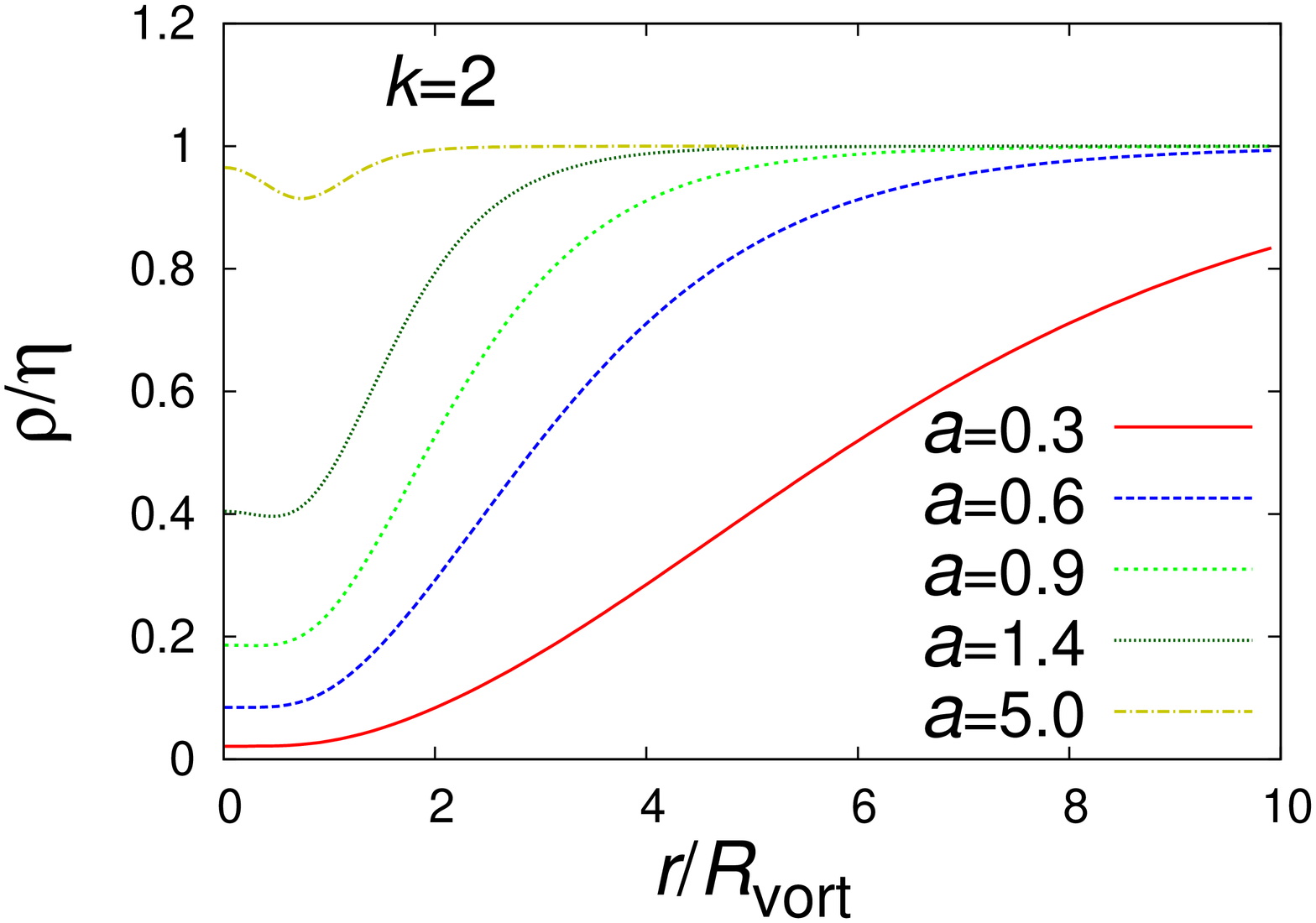}
  \includegraphics[width=55mm, angle=-90]{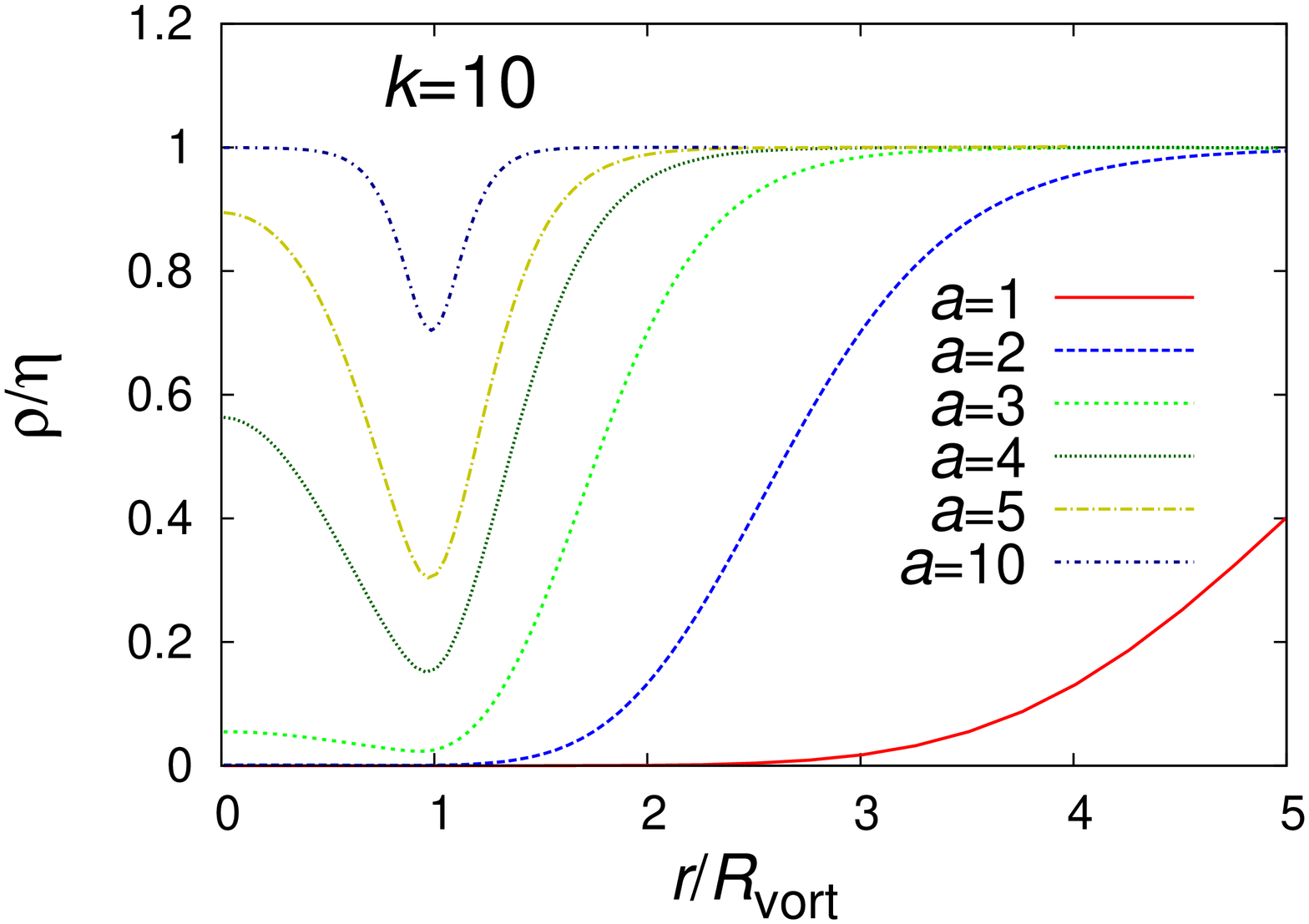}
\end{center}
\vskip -3mm
  \caption{ Condensate $\rho$ at the distance $r$ from the vortex center for
           various sizes (thicknesses) of the vortex cores $R_{\mathrm{vort}} = a/e \eta $.
           The vortices have vorticity numbers $k=1$ (top), $k=2$ (middle), and $k=10$ (bottom).}
  \label{fig:k1}
\end{figure}

The solutions of Eq.~\eq{eq:rho} can be found numerically. In Fig.~\ref{fig:k1} we show the condensate $\rho$, Eq.~\eq{eq:rho12}, for a vortex
carrying a single unit of the flux, $k=1$ (the plot at the top); a double-flux vortex, $k=2$ (the plot at the middle); and
the high-vorticity solution, $k=10$ (the plot at the bottom). One can observe that while the single-flux solutions $\rho = \rho(\xi)$
are always monotonic, the higher-flux solutions are not. A nonmonotonic density variation but for a different kind of vortices in a
two-component model was also found in Ref.~\cite{ref:nonmonotonic}. The solutions with a large-sized core ($a \gg 1$) and with a large
flux ($k=10$ in our example) have a well-recognized minimum around $r \approx R_{\mathrm{vort}}$.

\begin{figure}[!htb]
\begin{center}
  \includegraphics[width=55mm, angle=-90]{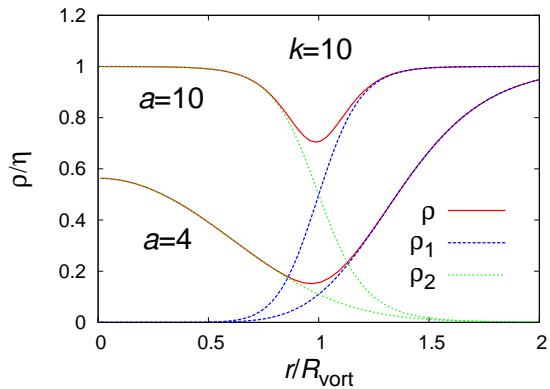}
\end{center}
\vskip -3mm
  \caption{ Condensates $\rho,$ $\rho_1$ and $\rho_2$ in
  high-vorticity ($k=10$) solutions with two
  sizes $a = e\eta R_{\mathrm{vort}}$ of the vortex cores.}
  \label{fig:k10:separate}
\end{figure}
The condensates for these vortex solutions are nonzero, $\rho \neq 0$, both at their centers, $r=0$,
and at spatial infinities, $\rho \to \infty$.
As for the individual condensates, the $\rho_1$ condensate is vanishing at the center, while the $\rho_2$ condensate is zero at spatial
infinity. This particular pattern of the condensates in the vortexlike solution is quite common for many-component systems. It appears
in particle physics in the context of the standard electroweak model~\cite{Achucarro:1999it}, and in the condensed matter physics: the corresponding
structures were experimentally observed in a two-component condensate of ${}^87$Rb atoms~\cite{ref:Matthews}.
We plot the individual condensates of two solutions with $a=4$ and $a=10$ for the $k=10$ vortex in Fig.~\ref{fig:k10:separate}.

\begin{figure}[!htb]
\begin{center}
\begin{tabular}{cc}
  \includegraphics[width=75mm, angle=0]{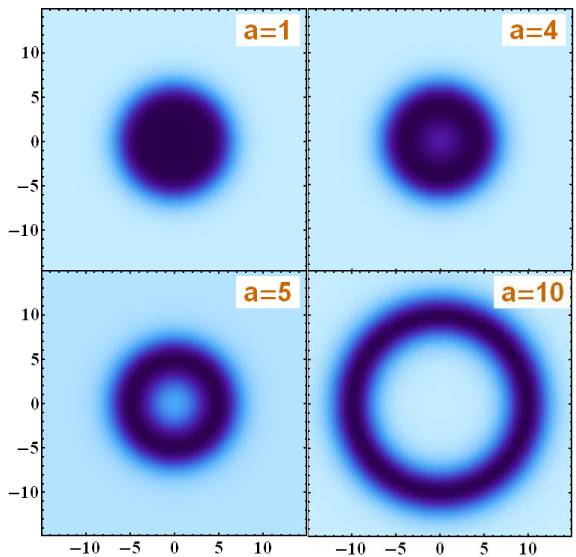}
  \end{tabular}
\end{center}
\vskip -3mm
  \caption{ Strength of the magnetic field~\eq{eq:B} in high\--vor\-ti\-ci\-ty ($k=10$) solutions
  in the transverse plane. Various sizes of the vortex cores $R_{\mathrm{vort}}=a/e\eta$ are shown. The
  spatial coordinates are given in units of~$e\eta$.}
  \label{fig:k10:2d}
\end{figure}

In general, the magnetic field~\eq{eq:B} of a vortex with a multiple vorticity is concentrated outside the center of the vortex.
In Fig.~\ref{fig:k10:2d} we show the density plots of the magnetic field
in the transverse ($x_1,x_2$) plane for $k=10$ vortices of different sizes $R_{\mathrm{vort}}$.
One can clearly see that the maximum of the magnetic field (darker regions) is distributed in a (generally) cylindrical
thin region of the radius $R \simeq R_{\mathrm{vort}} \equiv a/e \eta$.
Thus, the magnetic field of the vortex is concentrated, basically, within a thin pipe of the radius $R_{\mathrm{vort}}$.
The magnetic field inside and outside the pipe is zero. A possible existence of similar solutions with pipelike (``helical'')
structure in this model was discussed in~\cite{Niemi:2000ny}. The annular vortex solutions that somewhat resemble
our pipelike solutions were also discussed in Refs.~\cite{ref:Govaerts:2001,ref:Govaerts:2000}.

The pipelike structure is well pronounced for large ($a \gg 1$) vortices with high-vorticity numbers.
In a different context, vortices with high winding numbers were also studied in the standard electroweak
model in~\cite{Achucarro:1993bu}.

\begin{figure}[!htb]
\begin{center}
  \includegraphics[width=75mm, angle=0]{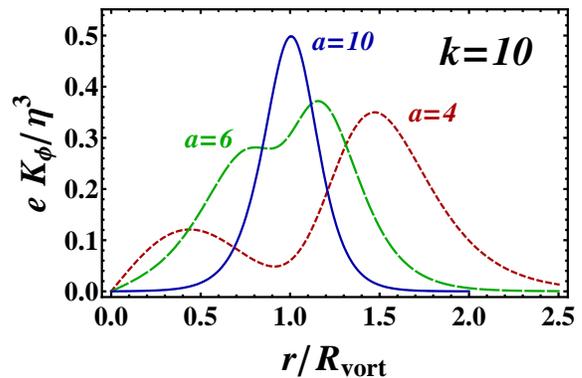}
\end{center}
\vskip -3mm
  \caption{ The spatially transverse diagonal isocurrent~\eq{eq:KT} for vortices of various sizes~$a$
  with a large winding number $k=10$.}
  \label{fig:K:phi}
\end{figure}

Let us now consider the spatially-transverse component of the conserved isovector current~\eq{eq:K}
which corresponds to the flow of the global $SU_I(2)$ charge. The diagonal (in the isospace) component
of this current is
\beqn
K^{(3)}_i(\xi;k,a) = - \frac{\varepsilon_{ij}}{a} \frac{x_i}{\xi} \frac{\partial}{\partial \xi}
\Bigl[\frac{\xi^{2k} -1}{\xi^{2k} + 1} \rho^2(\xi;k,a)\Bigr]\,.
\label{eq:KT}
\eeqn
In the polar coordinates of the transverse plane, the radial component of the transverse isocurrent~\eq{eq:KT} is zero, $K^{(3)}_r = 0$.
The angular component of this current is plotted in Fig.~\ref{fig:K:phi} for a few vortices with the large (fixed) winding number $k=10$.
The angular component exhibits a familiar pipelike structure with two peaks at moderate values of the vortex size $a$.
These peaks are merged into one peak at larger values of $a$.

\section{Longitudinal currents}
\label{sec:longitudinal}

The vortices may host certain waves that propagate freely along the vortices. The simplest way to introduce
such a wave is to generalize the meromorphic function $u$ [given, for a static solution, by Eq.~\eq{eq:sol}]
as follows (remember that $z=x_1 + i x_2$):
\begin{equation}
u \bigl(x_1,x_2, x_3, x_4\bigr) = \left(\frac{a}{z}\right)^{k} \! e^{i(\omega x_0 - p_3 x_3)}\,.
\label{eq:sol:wave}
\end{equation}
Here the wave frequency $\omega \equiv p_0$ and the wave momentum $p_3$ are two unknown parameters.
According to \eq{eq:rescaling}, in Eq.~\eq{eq:sol:wave} the wave frequency $\omega$ and the wave momentum $p_3$
are given in units of $e \eta$.

According to Eq.~\eq{stereodef} the ansatz~\eq{eq:sol:wave} corresponds (up to a gauge transformation)
to the following behavior of the scalar doublet~\eq{eq:Phi}:
\beqn
\Phi =
\left(
\begin{array}{l}
\rho_1(r) \, e^{i k \varphi} \\
\rho_2(r) \, e^{i(p_3 x_3 - \omega x_0)}
\end{array}
\right)\,.
\label{eq:Phi2}
\eeqn
Here $r$ and $\varphi$ are the polar coordinates in the transverse plane and $\rho_{1,2}$ are certain functions of $r$.
The form of Eq.~\eq{eq:Phi2} is essentially the same as in Refs.~\cite{Forgacs:2005sf} and \cite{Abraham:1992hv}.
The ansatz~\eq{eq:sol:wave} for the meromorphic function $u(z)$ was used in a different context in a $O(3)$ model
in Ref.~\cite{Ferreira:2008nn}.

In analogy with the previous section, we substitute the ansatz~\eq{eq:sol:wave} into the classical equations of motion,
\eq{eq1} and \eq{eq2} using Eq.~\eq{stereodef}. The new equations also include the electric density, $J_0$, and the
longitudinal electric current $J_3$, that are treated in these equations as the unknown independent functions. However,
one can immediately figure out that the second-order differential equations \eq{eq1} and \eq{eq2} are self-consistent provided that
\beqn
\omega =  \pm p_3\,, \qquad J_0(x) = \mp J_3(x) \equiv j \rho^2\,.
\label{eq:omega:j}
\eeqn

The first relation in Eq.~\eq{eq:omega:j} indicates that the longitudinal current is carried by massless waves. The second relation
means that the regions with nonzero electric current density are always electrically charged. The signs in Eq.~\eq{eq:omega:j}
distinguish between two possible directions of the wave propagation along the vortex.

The presence of the longitudinal currents does not influence the behavior of the condensate $\rho = \rho(r)$ because Eq.~\eq{eq:rho} is
valid in the $J_0 \neq 0$ case. The transverse components, $\mu=1,2$, of the electric current $J_\mu$ are not modified by the presence of the
longitudinal current as well, and Eq.~\eq{eq:JK} remains unaffected. As for the longitudinal components, the
(reduced) electric charge density $j \equiv J_0/\rho^2$, Eq.~\eq{eq:reduced}  -- or, equivalently, the longitudinal current
density, Eq.~\eq{eq:omega:j} -- is determined by the following differential relation:
\beqn
\left[\frac{\partial^2}{\partial \xi^2} + \frac{1}{\xi} \frac{\partial}{\partial \xi} - a^2 \rho^2(\xi)\right]
j(\xi)= \frac{4 \omega k^2 \xi^{2 k - 2}}{(\xi^{2 k}+1)^3} (\xi^{2 k} - 1). \qquad
\label{eq:j:eq}
\eeqn

\subsection{Longitudinal electric current}

The {\it total} longitudinal electric current $I_3^{\mathrm{tot}} \equiv \int {\mathrm{d}} x_1 {\mathrm{d}} x_2 J_3$
streaming through any transverse plane of the vortex, and the total electric charge
$\varrho^{\mathrm{tot}}\equiv \int {\mathrm{d}} x_1 {\mathrm{d}} x_2 J_3$ per unit vortex length
are both zero. Indeed, an integration of the left- and right-hand sides of Eq.~\eq{eq:j:eq}
over the spatial coordinates of the two-dimensional transverse plane gives us the vanishing result:
\beqn
\varrho^{\mathrm{tot}} = \pm I_3 \equiv \int {\mathrm{d}}^2 r \, j(r) \rho^2(r) = 0 \,.
\label{eq:Itot}
\eeqn
Here we used  the second equation in \eq{eq:omega:j} and noticed that $\xi \frac{\partial}{\partial \xi} j(\xi) \to 0$
in the limits $\xi \to 0$ and $\xi \to \infty$.

\begin{figure}[!htb]
\begin{center}
  \includegraphics[width=55mm, angle=-90]{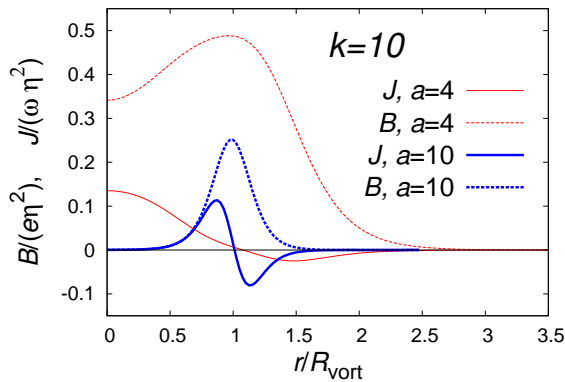}
\end{center}
\vskip -3mm
  \caption{ Magnetic field $B$ (the dashed lines) and the longitudinal electric current $J$ (the solid lines) in the $k=10$
  vortex as the function of the radius $r$. The red/thin (blue/thick) lines correspond to the vortex of the size $a=4$ ($a=10$).}
  \label{fig:k10:jB}
\end{figure}
Despite the total electric current flowing through any transverse plane is zero,
the local structure of the vortex in terms of electric currents is, however, nontrivial.
In Fig.~\ref{fig:k10:jB} we plot the density of the electric current and the strength of the
magnetic field as functions of the distance to the vortex with the vorticity $k=10$.
We show these quantities for two vortices with the sizes $a=4$ and $a=10$. In both cases
the electric current changes the sign at a certain nonzero distance $R_0$ from the center of
the vortex,
\beqn
J(R_0) = 0\,, \qquad 0 < R_0 <\infty\,.
\label{eq:R0}
\eeqn
For a typical vortex the alternation of the current takes place at the value of the radius, that is approximately
equal to the size of the vortex core, $R_0 \simeq R_{\mathrm{vort}}$. As one can see from Fig.~\ref{fig:k10:jB},
it is the very radius where the magnetic field takes its maximum. The density of the
current is positive in the inner core of the vortex, $r \lesssim R_{\mathrm{vort}}$
and negative in the outer space, $r \gtrsim R_{\mathrm{vort}}$.

\begin{figure}[!htb]
\begin{center}
  \includegraphics[width=55mm, angle=0]{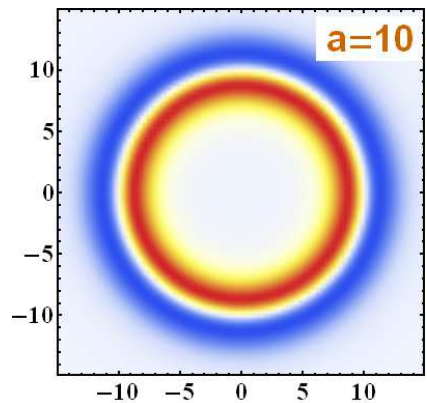}
\end{center}
\vskip -3mm
  \caption{ The density of the longitudinal current -- and, equivalently, the electric charge density --
  in the transverse plane of the $k=10$ vortex of the size $a=10$.
  The inner (red) pipe corresponds to the electric current flowing in the positive direction, while outer (blue) pipe
  denotes the density of the negatively directed current. There is almost no current inside the inner pipe and outside
  the outer pipe and also in between these two pipes.}
  \label{fig:k10:electric}
\end{figure}

In Fig.~\ref{fig:k10:electric} we show a two-dimensional plot of the distribution of the {\it electric} current in the transverse
plane of the $k=10$ vortex of the size $a=10$. The current is concentrated in the two nested narrow pipes that have
relatively large diameters. The electric current is almost zero outside these pipes. Notice, that due to Eq.~\eq{eq:omega:j}
the current-carrying pipes are also electrically charged.
The electrically charged pipes enclose the pipe of the magnetic field, which was visualized in the lower right plot of
Fig.~\ref{fig:k10:2d}. The interior and exterior of the vortex
are oppositely charged, as illustrated in Fig.~\ref{fig:k10:electric}.

The distribution of the transverse electric currents in the transverse plane have a qualitative similarity with the distribution
of the longitudinal currents, Fig.~\ref{fig:k10:2d}. The transverse currents -- defined by the first relation in Eq.~\eq{eq:JK} --
rotate in opposite directions inside and outside the pipe of the magnetic field. The transverse current density vanishes approximately
at the radius of the magnetic pipe, similarly to the density of the longitudinal currents.
These properties are the natural result of the nonmonotonic behavior of the condensate $\rho(r)$, Fig.~\ref{fig:k1}. Physically,
the outer current (say, circulating the vortex clockwise) generates the magnetic field along the $z$ axis in an ``outer'' tube of a certain radius.
The inner current (that rotates in the counterclockwise direction) leads to  appearance of the magnetic field in the opposite direction in the
``inner'' tube. Since the inner and outer tubes have different radii, the resulting profile of the magnetic field has a maximum at a certain
nonzero radius, thus forming a pipelike structure.

\begin{figure}[!htb]
\begin{center}
  \includegraphics[width=75mm, angle=0]{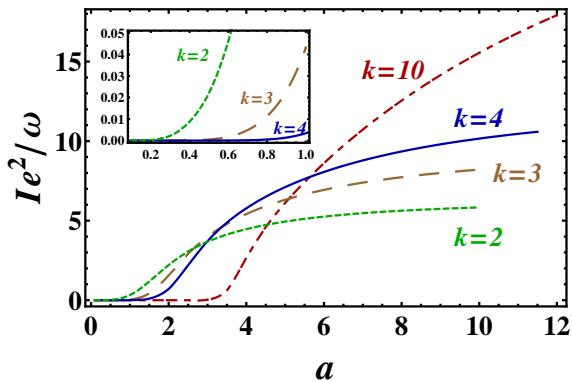}
\end{center}
\vskip -3mm
  \caption{ The net positive/negative current~\eq{eq:I} carried by interior/exterior
  of the vortex with the winding number $k$ vs the vortex size $a = e \eta R_{\mathrm{vort}}$.
  The inset shows zoom in on the region with $a < 1$.}
  \label{fig:I}
\end{figure}

The total positive and negative longitudinal current, $I^{(\pm)} = \pm I^{\mp}$, can be calculated by integration over,
say, the interior of the vortex core:
\beqn
I \equiv |I^{(\pm)}| = \int_{r \leqslant R_0} \dd^2 r \, J(r)\,,
\label{eq:I}
\eeqn
where we take, for definiteness, $J(r)>0$. The radius $R_0$ is defined as a nonzero finite distance
from the vortex center at which the electric current vanishes~\eq{eq:R0}. The integrated current~\eq{eq:I}
is shown in Fig.~\ref{fig:I} as a function of the vortex size $a$ for a few fixed vorticities.
For small vortices, $a \lesssim 1$, the total positive/negative current is
very small. However, the inner and outer pipes of the thick vortices, $a \gtrsim 1$ may carry rather strong electric current.
Moreover, the larger the transverse size of the vortex the stronger is the electric current~\cite{foot1}.

\begin{figure}[!htb]
\includegraphics[width=30mm, angle=-90]{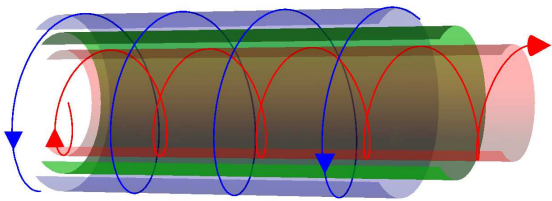}
  \caption{ A graphical representation of the structure of a thick ($a \gg 1$) vortex
  with a high winding number ($|k| \gg 1$). The pipelike magnetic field -- shown by the green surface in the middle -- is
  layered in between positive (inner) and negative (outer) electrically charged pipes (shown by the red and blue surfaces,
  respectively). These electric pipes carry helical electric currents in opposite directions. The arrows mark directions
  of the helical electric currents.}
  \label{fig:illustration}
\end{figure}

Summarizing this section, we visualize the internal structure of the pipelike vortex in Fig.~\ref{fig:illustration}.
The magnetic field has a pipelike profile shown by the green surface which is sandwiched
in between two other pipes. The inner and outer pipes carry the electric currents in positive (shown by the red surface)
and negative (the blue surface) directions, respectively. Since these electric currents in these pipes have both
longitudinal and transverse components, the three-dimensional structure of the electric currents resembles spirals
which are also visualized in Fig.~\ref{fig:illustration}.

\subsection{Longitudinal isocurrent}

Now let us now consider the longitudinal components of the conserved isovector current~\eq{eq:K}.
According to Eqs.~\eq{stereodef} and \eq{eq:sol:wave}, the off-diagonal components of the isocurrent, $K^{(1)}_\mu$
and $K^{(2)}_\mu$, are oscillating in time and space because they depend linearly on periodic functions
$\cos \vartheta$ and $\sin \vartheta$, where $\vartheta = \omega (x_0 \pm x_3) - \varphi$. The diagonal components are
time-independent quantities. The longitudinal isocurrent is
\beqn
K^{(3)}_3(\xi;k,a) & = & - 2 \omega \frac{\rho^2(\xi;k,a)}{\xi^{2k} + 1} \Bigl[\frac{\xi^{2k}}{\xi^{2k} + 1} \nonumber\\
& & + (1-\xi^{2k}) j(\xi;k,a){\Bigl|}_{\omega=1}\Bigr]\,.
\label{eq:KL}
\eeqn
The isocharge density, $K_0$, is linked with the longitudinal isocurrent density, $K_3$, in a manner of relation \eq{eq:omega:j} for the
electric charge/current density: ${\vec K}_3(x) = \pm {\vec K}_0(x)$. Both the isocharge density and the longitudinal
isocurrent are proportional to the frequency parameter~$\omega$.

\begin{figure}[!htb]
\begin{center}
  \includegraphics[width=75mm, angle=0]{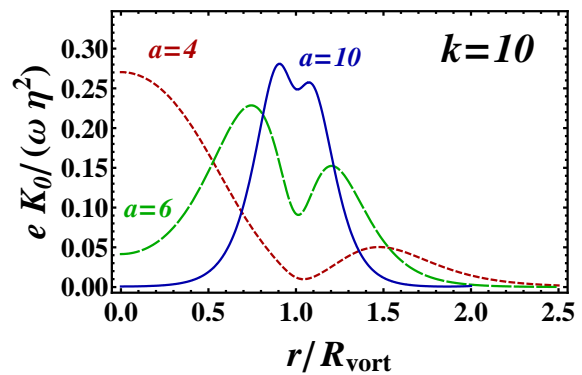}
\end{center}
\vskip -3mm
  \caption{ The longitudinal isocurrent~\eq{eq:KL} for vortices of various sizes~$a$
  with large winding number $k=10$.}
  \label{fig:K:0}
\end{figure}

The longitudinal isocurrent~\eq{eq:KL} is shown in Fig.~\ref{fig:K:0}. This current has a two-peak structure for all studied values
of the core size~$a$. The peaks are centered near the values of the radius that approximately correspond to the
extremes of the longitudinal electric current, plotted in Fig.~\ref{fig:k10:jB} for $a=4$ and $a=10$.

In addition, Fig.~\ref{fig:K:0} demonstrates that -- unlike the ordinary electric current -- the isocurrent density is either
positively {\it or} negatively
valued. Thus, the net isocurrent streaming along the vortex is nonvanishing and the vortex has a nonzero total isocharge. These global
properties are qualitatively similar to the characteristics of the solutions with low winding numbers that were
found in Refs.~\cite{Forgacs:2005sf} and \cite{Abraham:1992hv}.

\begin{figure}[!htb]
\begin{center}
  \includegraphics[width=75mm, angle=0]{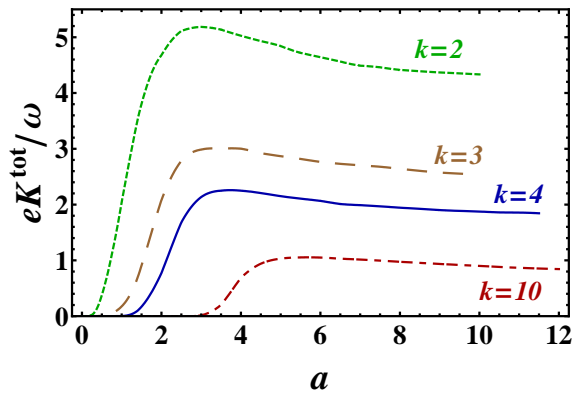}
\end{center}
\vskip -3mm
  \caption{ The net longitudinal isocurrent \eq{eq:Kfull}
  vs the vortex core size $a = e \eta R_{\mathrm{vort}}$ for various winding numbers $k$.}
  \label{fig:K:full}
\end{figure}

The absolute value of the total longitudinal isocurrent which is carried by the vortex,
\beqn
K^{\mathrm{tot}} = \left|\int \dd^2 r \, K_3^{(3)}(r)\right|\,,
\label{eq:Kfull}
\eeqn
is shown in Fig.~\ref{fig:K:full} as a function of the vortex core size $a$ for a few values of the winding number $k$.
The behavior of the new isocurrent~\eq{eq:Kfull} is similar to the net positive/negative electric current, Fig.~\ref{fig:I}.
The total isocurrent is very small (at fixed $\omega$) for small values of the vortex core size~$a$.
At a particular vortex size $a$, the isocurrent starts to grow quickly as a function of $a$. Then it reaches a maximum, and,
unlike the positive/negative fraction of the electric current, the isocurrent slowly diminishes as the function of the vortex
core size $a$. In general, the larger is the winding number $k$, the smaller is the net isocurrent.

Notice that the total longitudinal electric charge carried by the solutions vanishes exactly~\eq{eq:Itot} because this current is
generated by the {\it local} $U(1)$ symmetry~\eq{eq:Uem}. The net longitudinal isocurrent is not zero~\eq{eq:Kfull} since it
carries the charges corresponding to the global $SU(2)$ subgroup of the model~\eq{eq:SU2I}.

\section{Energy of vortices}
\label{sec:energy}

\subsection{Contribution of longitudinal current}

In this section we calculate the energy of the vortices that is carried by the longitudinal currents in the Bogomolny limit.
The energy momentum tensor of the Abelian two-Higgs model~\eq{lagrange} is given by the following equation,
\beqn
T_{\mu\nu} & = & - g_{\mu\nu} {\cal L} - F_{\mu}^{\sigma}F_{\nu\sigma}
\label{eq:EMT}\\
& & +\frac{1}{2}\left(D_{\mu}\Phi\right)^{\dagger}D_{\nu}\Phi+
\frac{1}{2}\left(D_{\nu}\Phi\right)^{\dagger}D_{\mu}\Phi\,,
\nonumber
\eeqn
where $g_{\mu\nu}$ is the metric tensor.
In the spherical isovector representation~\eq{eq:n} the energy density reads as follows:
\beqn
 T_{00} & =& \frac{1}{2}(\partial_{0}\rho)^{2}+\frac{1}{2}(\partial_{i}\rho)^{2}
            +\frac{\rho^{2}}{8}(\partial_{0}\vec{n})^{2} + \frac{\rho^{2}}{8}(\partial_{i}\vec{n})^{2}\nonumber\\
  & & +  \lambda(\rho^{2}-1)^{2} + \frac{\rho^{2}}{2}j_{0}^{2}+ \frac{\rho^{2}}{2}j_{i}^{2} \label{eq:T00} \\
  & & + \frac{1}{2}\left(j_{0i}-\frac{1}{2}f_{0i}\right)^{2} + \frac{1}{4}\left(j_{ij}-\frac{1}{2}f_{ij}\right)^{2}\,.
\nonumber
\eeqn
We integrate Eq.~\eq{eq:T00} over the transverse coordinates $x_1$ and $x_2$,
imply the Bogomolny limit~\eq{eq:Bogomolny}, and simplify the expression using both integration by parts and
the corresponding equations of motion~\eq{eq:bogomolny1} and \eq{eq:bogomolny1}. As a result, we get the
vortex energy per unit vortex length:
\beqn
\sigma(k,a)
\equiv
\int {\mathrm{d}}^2 x\, T_{00}
& =& \pi |k| + \chi(k,a) \, \omega^2\,.
\label{string_tension}
\eeqn

The first term in the right-hand side of Eq.~\eq{string_tension} corresponds to the energy of the vortex in the absence of the
electric current~\eq{eq:energy:functional}. The second part is the contribution of the electric current to the vortex energy.
This contribution is proportional to the wave frequency square, $\omega^2$ or, equivalently, to the square of the total
isocurrent carried by the vortex~\eq{eq:Kfull}, or to the total electric current in, say, the inner charged pipe of the solution~\eq{eq:I}.

The dependence of the energy on the current is governed by the dimensionless coefficient $\chi$ that can be separated into
three parts:
\beqn
\chi(k,a) & = & \chi_{1}(k,a) + \chi_{2}(k,a) + \chi_{3}(k) \,.
\label{eq:chi}
\eeqn
The first term in Eq.~\eq{eq:chi} is expressed via the condensate~$\rho$,
\beqn
\chi_1(k,a) & =& 2 \pi a^2 \int\limits_0^\infty\dd \xi \,\frac{\xi^{2k+1} \rho^{2}(\xi;k,a)}{(\xi^{2k}+1)^{2}} \,,
\label{eq:chi1}
\eeqn
while the second term can be determined via the electric current density~$j$
\beqn
\chi_{2}(k,a) & =&  8 \pi k^2 \int\limits_0^\infty \dd \xi \, \frac{(\xi^{2k}- 1) \, \xi^{2k-1}}{(\xi^{2k} + 1)^{3}}
j(\xi;k,a){\Bigl|}_{\omega=1} \!,\quad\,.
\label{eq:chi2}
\eeqn
The third term can be calculated exactly:
\beqn
\chi_{3}(k) & =& 8 \pi k^2 \int\limits_0^\infty \dd \xi \, \frac{\xi^{4k-1}}{(\xi^{2k}+1)^{4}}
\equiv \frac{2\pi |k|}{3} \,.
\label{eq:chi3}
\eeqn
The energy of the vortex~\eq{string_tension} saturates the Bogomolny bound and it thus corresponds to
the lowest possible energy for fixed parameters~$k$, $\omega$ and $a$. Expressions~\eq{string_tension}--\eq{eq:chi3}
correspond to the lower bound on energy calculated previously in Ref.~\cite{Forgacs:2005sf}.

The first integral~\eq{eq:chi1} diverges for $k=1$. Indeed, at large transverse distances $\xi \gg 1$ the condensate is
distance independent, $\rho \simeq 1$, and the integral in Eq.~\eq{eq:chi1} becomes logarithmically divergent at large distances,
$\int^{\xi_{\mathrm{\max}}} \xi^{-1}\dd \xi \sim \log \xi_{{\mathrm{\max}}}$.
Thus, finite-sized vortices ($a \neq 0$) with a minimal winding number, $k=1$, cannot carry
the longitudinal currents in the Bogomolny limit. However, the vortices with higher vorticities, $k \geqslant 2$, can support
the longitudinal currents in the this limit.

\begin{figure}[!htb]
\begin{center}
  \includegraphics[width=75mm, angle=0]{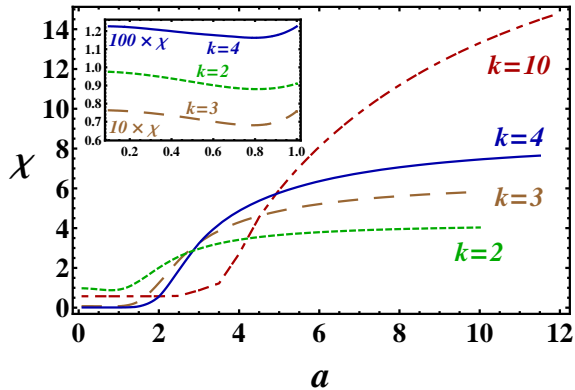}
\end{center}
\vskip -3mm
  \caption{ The coefficient $\chi$ that determines the contribution of the electric current into the energy of
  the vortex~\eq{string_tension}.  The inset shows a zoom in on a low-$a$ region.}
  \label{fig:chi}
\end{figure}

In the Bogomolny limit the energy of the currentless vortex -- given by the first term in Eq.~\eq{string_tension} --
is independent of the vortex size $a$. Unexpectedly, in the presence of a longitudinal current the energy becomes dependent
on the vortex size parameter~$a$.
In Fig.~\ref{fig:chi} we plot $\chi$ as a function of the vortex size $a$ for the same set of vorticities that was already
selected for Fig.~\ref{fig:I}.
Notice that despite the fact that the curves in Fig.~\ref{fig:I} and in Fig.~\ref{fig:chi} look
similar on a qualitative level, they cannot be superimposed on each other by simple scale transformations.

According to Fig.~\ref{fig:chi}, the excess of energy provided by the
electric current is a monotonically increasing function of the vortex size at relatively large size $a$  of the vortex core.
A zoom in on the low-$a$ region  -- shown in the inset of Fig.~\ref{fig:chi} -- reveals a very shallow minimum
at the fixed size of the vortex
\beqn
a_{\mathrm{min}} \approx 0.8\,, \quad \mbox{or} \quad R_{\mathrm{vort}} \approx \frac{0.8}{e\eta}\,,
\label{eq:a:min}
\eeqn
for all studied vorticities $k \geqslant 2$.

Thus, in the Bogomolny limit a vortex with a fixed winding number $k>1$ and a fixed frequency parameter $\omega \neq 0$
would tend to change its transverse size to the value~\eq{eq:a:min} that corresponds to the global minimum of the vortex energy.
According to the inset of Fig.~\ref{fig:I}, the electric current at $a \simeq a_{\mathrm{min}}$ is small but nonzero (we remind that
the current can be made arbitrarily large by increasing the free parameter $\omega$).

The vortex core tends to stabilize its transverse size at the value~\eq{eq:a:min} for which the nested-pipe structure of the vortex interior
cannot be resolved visually. This fact is seen in the behavior of the condensate $\rho$ of the $a=1$ vortex with the vorticity $k=10$
(lowest panel of Fig.~\ref{fig:k1}), or one can alternatively look at the magnetic field profile of the same vortex in Fig.~\ref{fig:k10:2d}.
However, the minimum is very shallow so that a small perturbation -- for example, induced by a shift of parameters towards
type-II superconductivity -- may shift the stability point to higher values of the vortex core radius~$a$.
\begin{figure}[!htb]
\begin{center}
  \includegraphics[width=75mm, angle=0]{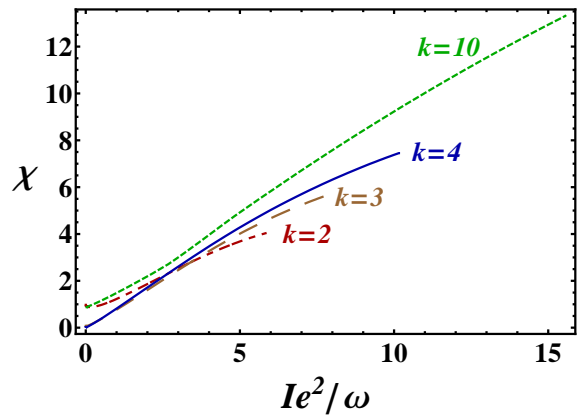}
\end{center}
\vskip -3mm
  \caption{ The energy coefficient $\chi$ as a function of the (normalized) absolute value of the electric
  current flowing in, say, the inner charged pipe~\eq{eq:I}.}
  \label{fig:Current:Energy}
\end{figure}

The dependence of the vortex energy per unit vortex length on the value of the electric current in one of
the pipes $I$ is shown in Fig.~\ref{fig:Current:Energy}. At very low values of the current $I$ the energy has a
very shallow minimum which is visible in the plot of the $k=2$ solution. At higher values of the current
the energy is a monotonically rising function.

\subsection{Rearrangement of vortex lattice}

As we have already discussed, in the Bogomolny limit the energy per unit vortex length of an elementary ($k=1$) vortex is infinite
at any finite $\omega$, or, equivalently, at any nonzero value of the longitudinal isocurrent. In the absence of the external isocurrent
the total energy of a many-vortex configuration does not depend neither on positions of vortices nor on their sizes in this limit.
Now, in a thought experiment, let us apply an external isocurrent $K^{\mathrm{tot}}$ along
a configuration of an even number of parallel elementary vortices. Because of energy considerations it is clear
that the elementary vortices would tend to merge together and form a set of, at least, double ($k=2$) vortices that have
a finite energy at the nonzero isocurrent. Then each of the $k=2$ vortices would tend to adjust its transverse size $a$
in such a way that the total energy is minimal at the fixed value of the isocurrent~\cite{foot2}.
Thus, the presence of an external isocurrent should generally lead to rearrangement of the vortex ensembles.

A nonzero isocurrent~\eq{eq:KL} corresponds to a combination of two electric currents
-- carried by up and down com\-po\-nents of the order parameter -- for which the total electric current is zero.
Although a realization of such an isocurrent in an experimental setup is an open issue, our result suggests that the vortex lattice
should be restructured under the influence of the external isocurrent that is parallel to the external magnetic field.
Also, our result shows that the multiple merging of the elementary vortices under the influence of the external isocurrent may lead
to a formation of a vortex with a high winding number that exhibits the exotic ``nested-pipe'' structure.

We expect that our results are rather generic and they should not be limited strictly to the particular limit~\eq{eq:Bogomolny}
of the couplings. If the system is sufficiently close to the Bogomolny limit, then we expect that an external isocurrent
would induce a rearrangement of the vortices in this system.

One of the well-studied examples of the two-band superconductor is MgB${}_2$~\cite{ref:MgB2:1}.
In external magnetic field the vortex lattice in MgB${}_2$ has quite specific geometrical patterns~\cite{ref:MgB2:2},
that are imprinted by elementary vortices. Such an inhomogeneous distribution of vortices in multicomponent systems
was first predicted theoretically in Ref.~\cite{ref:inhomogeneous}.
It would be interesting to check the effect of the external longitudinal isocurrent on the properties of the vortex
lattice in this real material, which is rather far from the Bogomolny limit~\eq{eq:Bogomolny}.

\section{Conclusion}

We found pipelike vortex structures in an Abelian gauge model with two scalar condensates. This model possesses global $SU_I(2)$ symmetry
(isosymmetry) as well as local $U_{e.m.}(1)$ symmetry corresponding to the Maxwell electromagnetism. The symmetry group is broken spontaneously
to a global $U(1)$ subgroup. The model possesses vortices that are linelike topological defects which carry magnetic flux. We found that
the vortices with high winding numbers and with large sizes of the vortex cores have the pipelike shape of the magnetic field:
the magnetic field is concentrated at a certain distance from the geometric center of the vortex, thus resembling a pipe.

For the sake of simplicity we were working in the Bogomolny limit of the model in which the classical equations of motion are drastically simplified.

We show that nonelementary vortices in the Bogomolny limit of the model are able to support the longitudinal isocurrents that carry
the isocharge along the vortices. The isocharge corresponds to the conserved Noether charge with respect to the diagonal component of
the global $SU_I(2)$ subgroup. The isocurrent is proportional to the difference between the electric currents carried by the upper and lower
components of the order parameter. The current-carrying vortex also possesses a nonzero density of the
isocharge localized in the vicinity of the vortex core. The elementary current-carrying vortices with unit vorticity have infinite energy
per unit length, while the energies of the current-carrying vortices with multiple winding numbers are finite.

The total vortex energy is given by a sum of two terms. The first contribution is the standard Bogomolny term that is proportional
to the magnetic flux inside the vortex. The second term comes due to the presence of the longitudinal current. This term is proportional
to the total isocurrent squared. The dependence of the vortex energy on the isocharge current may lead to potentially observable effects like,
for example, rearrangement of the positions of vortices in the vortex lattice of a two-band superconductor.

The vortices with large transverse sizes and with high winding numbers have a nested pipelike structure.
In addition to the mentioned ``magnetic pipe'' that carries the
magnetic flux of the vortex, the vortex also possesses two electrically charged pipes of larger and smaller radii. These ``electric pipes''
are oppositely charged and they also carry longitudinal electric currents of equal strength in opposite directions so that
the net electromagnetic current flowing through a transverse section of the vortex is always zero.
The magnetic pipe is always layered between the two electric pipes.

\vskip 3mm

The authors are grateful to P.~Forgacs, A.~Niemi, M.~Volkov and G.~Volovik  for interesting discussions, useful comments, and suggestions.
This work has been supported by the French Agence Nationale de la Recherche project ANR-09-JCJC ``HYPERMAG'' and
by the Swedish STINT Institutional Grant No. IG2004-2 025. M.N.Ch. is grateful to the members of the Department of
Theoretical Physics at Uppsala University for the kind hospitality and stimulating environment.


\begin{thebibliography}{99}
\bibitem{foot1}
Here we fix the value of the frequency parameter $\omega$ because of the proportionality law $I \propto \omega$ for the total current~\eq{eq:I}.
We also remind a reader that the electric currents in inner and outer pipes flow in opposite directions. The total
electric current streaming along the vortex is always zero, Eq.~\eq{eq:Itot}.
\bibitem{foot2}
The variation of the vortex size at a fixed value of the isocurrent $K^{\mathrm{tot}}(a,\omega) = \omega K^{\mathrm{tot}}(a,\omega=1)$
is realizable because of the presence of the free parameter $\omega$. The parameter $\omega$ is to be adjusted properly to compensate
variation in the current, Fig.~\ref{fig:K:full}, caused by the variation in the vortex size $a$.
\end{thebibliography}
\end{document}